\documentclass[11pt]{llncs}

\input{psfig.sty}
\usepackage{amsfonts,amssymb}
\newcommand{\mod}[1]{\mbox{ mod}\;#1}

\begin{document}

\title{Efficient Authenticated Encryption Schemes with Public Verifiability}

\author{Guilin Wang\inst{1}, Feng Bao\inst{1}, Changshe Ma\inst{2}, and Kefei Chen\inst{2}}
\institute{Institute for Infocomm Research (I$^2$R)\\
21 Heng Mui Keng Terrace, Singapore 119613 \\
\email{\{glwang, baofeng\}@i2r.a-star.edu.sg} \\
\and 
Department of Computer Science and Engineering \\
Shanghai Jiaotong University, Shanghai 200030, China\\
\email{\{mcs,kfchen\}@sjtu.edu.cn}} 

\maketitle

\begin{abstract}
An authenticated encryption scheme allows messages to be encrypted and authenticated simultaneously. In 2003, Ma and Chen proposed such a scheme with public verifiability. That is, in their scheme the receiver can efficiently prove to a third party that a message is indeed originated from a specific sender. In this paper, we first identify two security weaknesses in the Ma-Chen authenticated encryption scheme. Then, based on the Schnorr signature, we proposed an efficient and secure improved scheme such that all the desired security requirements are satisfied. \\

\end{abstract}

\section{Introduction}

In the areas of computer communications and electronic transactions, one of important topic is how to send data in a confidential and authenticated way. Usually, the confidentiality of delivered data is provided by encryption algorithms, and the authentication of messages is guaranteed by digital signatures. In the traditional paradigm, these two cryptographic operations are performed in the order of encrypt-then-sign. In \cite{HMP94}, Horster et al. proposed an efficient scheme such that messages can be encrypted and authenticated simultaneously. Later, Lee and Chang \cite{LC95} improved Horster et al.'s authenticated encryption scheme so that no hash function is needed. However, both schemes does not provide the property of not-repudiation, i.e., the receiver cannot prove to a third party that some messages are indeed originated from a specific sender. In \cite{Zhe97}, Zheng introduced signcryption schemes such that unforgeablility, confidentiality, and not-repudiation can be provided {\it simultaneously}. Since the non-repudiation protocols in \cite{Zhe97a} are based on zero-knowledge proofs, Zheng's schemes are inefficient when there are disputes between the receiver and the sender. 

In \cite{MC03}, Ma and Chen proposed an efficient authenticated encryption scheme with public verifiability. That is, in their scheme the receiver Bob can efficiently prove to a third party that a message is indeed originated from the sender Alice. However, this paper shows that their scheme is insecure since dishonest Bob can forge a valid ciphertext so that it looks as if it were generated by Alice. In our attack, the only assumption is that Bob registers his public key with a certification authority (CA) after he knows Alice's public key. This assumption almost always holds in the existing public key infrastructures (PKIs). Another problem in their scheme is that their public verification protocol does not work due to a mathematical error. To overcome these weaknesses in the Ma-Chen scheme, we propose a new scheme based on the Schnorr signature. In addition, technical discussions are provided to show that our scheme is secure and efficient. 

The rest of this paper is organized as follows. Section 2 first reviews the Ma-Chen scheme. Then, the security analysis is presented in Section 3. After that, we propose a new scheme and analyze it in Section 4. Finally, the conclusion is given in Section 5.

\section{Review of the Ma-Chen Scheme}

A third trusted party (TTP) selects a triple $(p,q,g)$, where $p$ and $q$ are two large primes satisfying $q|(p-1)$, and $g\in \mathbb{Z}_p^*$ is an element of order $q$. It is assumed that the Decisional Diffie-Hellman (DDH) problem are difficult in the cyclic group $G_q=\langle g\rangle$. That is, given $g, g^a, g^b, g^c\in G_q$ where $a$, $b$ and $c$ are unknown random numbers, it is infeasible to determine whether $g^{ab}\mod p$ equals $g^{c} \mod p$ \footnote{There are another two related numerical assumptions, i.e., the discrete logarithm (DL) problem and the computational Diffie-Hellman (CDH) problem are difficult in the cyclic group $G_q=\langle g\rangle$. That is, given $g, g^a, g^b\in G_q$ where $a$ and $b$ are unknown random numbers, it is infeasible to compute $a$ or $g^{ab}\mod p$. Actually, it is easy to know that DDH assumption is at least as strong as CDH assumption, and CDH assumption is at least as strong as DL assumption.}. In addition, the TTP publishes a secure hash function $H(\cdot)$. We assume that Alice and Bob have the certified secret/public key pairs $(x_A,y_A=g^{x_A}\mod p)$ and $(x_B,y_B=g^{x_B}\mod p)$, respectively. \vskip 1mm

To send a message $m\in {\mathbb Z}_p^*$ to the receiver Bob, the sender Alice does as follows.
\begin{description}
\item [(A-1)] Pick a random number $k\in \mathbb Z_q^*$, and then compute $v=(g\cdot y_B)^k \mod p$,  $e=v \mod q$.
\item [(A-2)] Set $c=m\cdot H(v)^{-1} \mod p$, $r=H(e,H(m))$, and $s=k-x_A\cdot r \mod q$.
\item [(A-3)] Send the triple $(c,r,s)$ to Bob via a public channel. 
\end{description}

Upon receiving $(c,r,s)$ from Alice, the receiver Bob does the following:
\begin{description}
\item [(B-1)] Compute $v=(g\cdot y_B)^s\cdot y_A^{r(x_B+1)} \mod p$ and $e=v \mod q$.
\item [(B-2)] Recover the message $m=c\cdot H(v) \mod p$, and check whether $r\equiv H(e,H(m))$. 
\item [(B-3)] If $r\equiv H(e,H(m))$ holds, Bob concludes that $(c,r,s)$ is indeed encrypted by Alice. 
\end{description}

For public verification, Bob first computes $K_1=(y_B^s \cdot y_A^{r\cdot x_B} \mod p)\mod q$, and then forwards $(H(m),K_1,r,s)$ to the arbitrator TTP. The TTP performs as follows: 
\begin{description}
\item [(TTP-1)]\quad Compute $e'=(g^s\cdot y_A^r\cdot K_1\mod p)\mod q$.
\item [(TTP-2)]\quad If $r\equiv H(e',H(m))$, the TTP knows that Alice is the originator of encryption and signature. 
\end{description}

\section{Security Analysis of the Ma-Chen Scheme}

We now give some explanations and remarks on the Ma-Chen scheme. Note that to decrypt and verify the triple $(c,r,s)$, one needs to know the value of $x_B$ or $y_{AB}$, where $y_{AB}=g^{x_A\cdot x_B} \mod p$. Actually, using $y_{AB}$ the ciphertext $(c,r,s)$ can be easily decrypted and verified as follows: (a) compute $v=(g\cdot y_B)^s \cdot y_{AB}^{r}\cdot y_A^r\mod p$, and $e=v \mod q$; (b) recover $m=c\cdot H(v)\mod p$, and check whether $r\equiv H(e,H(m))$. Therefore, the value of $y_{AB}$ can not be revealed to anybody other than Alice and Bob. The authors of \cite{MC03} noticed this problem. Therefore, in their scheme only the tuple $(H(m),K_1,r,s)$ (not including the value of $v$) is revealed to the TTP, so that even the TTP cannot derive the value of $y_{AB}$. The reason is that if one value of $v$ is known by the TTP (or anybody else), then $y_{AB}$ can be obtained easily by $y_{AB}=y_A^{-1}\cdot v^{r^{-1}}\cdot (g\cdot y_B)^{-s\cdot r^{-1}}\mod p$. In addition, recall that anybody cannot derive the value of $y_{AB}$ directly from Alice's public key $y_A$ and Bob's public key $y_B$, since it has been assumed that the DDH assumption holds in the multiplicative cyclic subgroup $G_q=\langle g\rangle$. 

Based on the above observations and the fact that the value of $s$ in the Ma-Chen scheme is computed in a very similar way as in the secure Schnorr signature scheme \cite{Sch91}, the authors of \cite{MC03} analyzed the security of their scheme, and claimed that their scheme satisfies the following security properties:
\begin{itemize}
\item [(1)] {\it Unforgeability}: Except Alice, any attacker (including Bob) cannot generate a valid ciphertext $(c,r,s)$ for a message $m$ so that the verification procedure (B-1 to B-3) or the the public verification procedure (TTP-1 to TTP-2) is satisfied.
\item [(2)] {\it Confidentiality}: Under the DDH assumption, any third party cannot derive the message $m$ from the ciphertext $(c,r,s)$. 
\item [(3)] {\it Non-repudiation}: Once Bob reveals $(H(m),K_1,r,s)$, anybody can verify that $(r,s)$ is Alice's signature. Therefore, the TTP can settle possible disputes between Alice and Bob. 
\end{itemize}

However, in the following we show that their scheme is forgeable actually. Moreover, we identify an design error in their public verification procedure. That is, even all parties follow the specifications of their scheme honestly, the TTP cannot conclude that $(H(m), K_1,r,s)$ is generated by Alice. Therefore, the Ma-Chen scheme does not meet the properties of unforgeability, non-repudiation and public verifiability. \vskip 1mm

{\bf Forgeability}. Firstly, the authors observed that Bob is the strongest attacker to forge a triple $(c,r,s)$, since he knows $x_B$ which is used in verification procedure. Then, they claimed that their scheme is {\it equivalent} to the Schnorr signature \cite{Sch91}. So they concluded that their scheme is unforgeable against adaptive attacks, as the Schnorr signature is proved to be unforgeable (in the random oracle model) \cite{PS00}. Unfortunately, we notice that this is not the fact, though the value of $s$ is indeed calculated in a very similar way as in the Schnorr signature \cite{Sch91}. To show this fact directly, we now demonstrate a concrete attack on the Ma-Chen scheme. In our attack, we assume that Bob registers his public key $y_B$ with a certification authority (CA) after he knows Alice's public key $y_A$. This assumption almost always holds in the existing public key infrastructures (PKIs). Anyway, in the original paper \cite{MC03}, it is not specified that Alice and Bob have to register their public simultaneously. Moreover, in many scenarios it seems worth to register or update a (new) public key in the point view of the (malicious) verifier Bob, even if such an action can only enable him to forge one valid ciphertex for one message. \vskip 1mm

To amount this attack, the verifier Bob forges a valid ciphertext $(c,r,s)$ for a message $m$ of his choice as follows. 
\begin{itemize}
\item [(1)] Pick two random numbers $a, b\in \mathbb Z_q^*$, and compute $v=g^a\cdot y_A^b \mod p$.
\item [(2)] Compute $e=v \mod q$, $c=m\cdot H(v)^{-1} \mod p$, $r=H(e,H(m))$, and $s=rab^{-1}\mod q$.
\item [(3)] Set his secret key $x_B=br^{-1}-1 \mod q$, and then register the public key $y_B=g^{x_B}\mod p$ with a certification authority (CA). 
\end{itemize}

We now show that the forged triple $(c,r,s)$ is a valid ciphertext for message $m$ with respect to the public keys $y_A$ and $y_B$. Firstly, the following equalities hold:
$$\begin{array}{lcl} (g\cdot y_B)^s\cdot y_A^{r(x_B+1)}\mod p & \equiv & g^{(1+x_B)s} \cdot y_A^b\mod p\\ & \equiv & g^a \cdot y_A^b\mod p \\ & \equiv & v. \end{array}$$
Then, we have $e\equiv v\mod q$, $m \equiv c\cdot H(v) \mod q$, and $r\equiv H(e,H(m))$. So $(c,r,s)$ is a valid triple. \vskip 1mm

In addition, note that even if Alice realizes later a triple $(c,r,s)$ was forged via the above attacking procedure, she cannot defence for herself by computing Bob's secret $x_B$. We explain the reasons as follows. Since the hash function $H(\cdot)$ is usually modelled as a random function, to derive the secret key $x_B$ the useful information for Alice is the following three equations: 
 $v=g^a\cdot y_A^b \mod p$, $s=rab^{-1}\mod q$, and $x_B=br^{-1}-1 \mod q$. Firstly, note that Alice (with her secret key $x_A$) cannot derive the values of $a$ and $b$ from equation $v=g^a\cdot y_A^b \mod p$, as there are $q-1$ candidates of such pairs $(a,b)$. More specifically, for any given $a\in Z_q^*$, there is a fixed value of $b$ such that $v=g^a\cdot y_A^b \mod p$. Secondly, Alice can try to derive the value of $x_B$ by eliminating $a$ and $b$ in the above three equations. To do so, she knows that $a=s(1+x_B)\mod q$ and $b=r(1+x_B) \mod q$. Consequently, Alice gets equation $v=(g^s\cdot y_A^r)^{(1+x_B)} \mod p$. To get the value of $x_B$ from this equation, Alice has to face the difficult discrete logarithm problem, which is widely believed intractable. \vskip 2mm

{\bf Design Error}.  We note that the TTP cannot validate a valid tuple $(H(m),K_1,r,s)$ by the public verification procedure, i.e., TTP-1 to TTP-2. The reason is that $e'\neq e$ even if Alice, Bob, and the TTP all are honest. Namely,   
$$[(g\cdot y_B)^s\cdot y_A^{r(x_B+1)} \mod p] \mod q \neq [g^s\cdot y_A^r\cdot K_1\mod p ] \mod q,$$
where $K_1=(y_B^s \cdot y_A^{r\cdot x_B} \mod p)\mod q$, $p$ and $q$ are two primes such that $q|(p-1)$. In the original paper \cite{MC03}, however, the those two expressions are considered as equivalent. This problem is also identified independently by Wen et al. \cite{WLH03}. For more details, please refer to their paper. 

\section{Improved Scheme}

In this section, we propose an improvement of the Ma-Chen scheme by exploiting the similar idea of Bao and Deng \cite{BD98}. However, different from Bao and Deng's work, we use the provably secure Schnorr signature as the underlying signature scheme. Furthermore, technical discussions are provided to show that our scheme is secure and efficient. 

\subsection{Description of the Scheme}

In our scheme, we assume that $(E_K(\cdot), D_K(\cdot))$ is a pair of ideal symmetric key encryption/decryption algorithms under the session key $K$. In addition, $h(\cdot)$ is another suitable hash function, which maps a number of $Z_p$ to a session key for our symmetric key encryption/decryption algorithms. Other notations are the same as in Section II. \vskip 1mm

To send a message $m\in {\mathbb Z}_p^*$ to the receiver Bob in an authenticated and encrypted way, the sender Alice does as follows.
\begin{description}
\item [(A-1)] Pick a random number $k\in \mathbb Z_q^*$, and then compute $t_1=g^k \mod p$, and $t_2=y_B^k \mod p$.
\item [(A-2)] Set $c=E_{h(t_2)}(m)$, $r=H(m,t_1)$, and $s=k+r\cdot x_A\mod q$.
\item [(A-3)] Send the triple $(c,r,s)$ to Bob via a public channel. 
\end{description}

Upon receiving $(c,r,s)$ from Alice, the receiver Bob does the followings:
\begin{description}
\item [(B-1)] Compute $t_1=g^s y_A^{-r} \mod p$, and $t_2=t_1^{x_B} \mod p$.
\item [(B-2)] Recovers the message $m=D_{h(t_2)}(c)$, and check whether $r\equiv H(m,t_1)$. 
\item [(B-3)] If $r\equiv H(m,t_1)$ holds, Bob concludes that $(c,r,s)$ is indeed encrypted by Alice. 
\end{description}

For public verification, Bob just needs to release $(m,r,s)$. Then, any verifier can check whether $(r,s)$ is a standard Schnorr signature for message $m$ as follows: 
\begin{description}
\item [(V-1)] Compute $t_1=g^s y_A^{-r} \mod p$.
\item [(V-2)] $(r,s)$ is Bob's valid signature for message $m$ if and only if $r\equiv H(m,t_1)$. 
\end{description}

\subsection{Security}

Our scheme is very simple in the logic structure. That is, we exactly use the Schnorr signature scheme to generate the pair $(r,s)$, i.e., the standard Schnorr signature for a message $m$. Therefore, according the provable security of the Schnorr signature given in \cite{PS00}, any adaptive attacker (including Bob) cannot forge a valid ciphertext $(c,r,s)$ for any message $m$ such that $m=D_{h(t_2)}(c)$ and $r\equiv H(m,t_1)$, where $t_1=g^s y_A^{-r} \mod p$, $t_2=t_1^{x_B} \mod p$. Otherwise, this implies that the attacker has successfully forged a valid Schnorr signature $(r,s)$ for a message $m$, which is in turn contrary to the provable security of the Schnorr signature scheme. So our scheme satisfied the unforgeability. 

Now we discuss the confidentiality, i.e., except the receiver Bob, anyone else cannot extract the plaintext $m$ from the ciphertext $(c,r,s)$. Firstly, note that an attacker cannot extract the plaintext $m$ from the equality $r\equiv H(m,t_1)$ after recovering $t_1=g^s y_A^{-r} \mod p$, since the secure hash function $H(m,t_1)$ hides the information of $m$. Another way for getting the message $m$ is to decrypt the ciphertext $c$ directly. To do so, the attacker has to obtain the session key $h(t_2)$ since $(E_K(\cdot), D_K(\cdot))$ is assumed to be an ideal (so secure) symmetric key encryption/decryption algorithm pair. This means that to get the session key $h(t_2)$, the attacker has to get the value of $t_2$ first, since $h(\cdot)$ is also a secure hash function. However, the attacker cannot get the value of $t_2$ from the values $t_1$ and $y_B$. In fact, the latter problem is the CDH problem, which is widely believed intractable in security community. Therefore, we conclude that our scheme meets the the confidentiality. 

Finally, the property of non-repudiation is also satisfied in our scheme due to the following two facts: (a) Only Alice can generate a valid ciphertext $(c,r,s)$; and (b) Anybody can very that $(r,s)$ is a standard Schnorr signature if the receiver Bob releases the triple $(m,r,s)$. Consequently, a TTP can easily settle potential disputes between Alice and Bob by checking whether $r\equiv H(m,t_1)$, where $t_1$ is computed by $t_1=g^s y_A^{-r} \mod p$. 

\subsection{Efficiency}

In our scheme and the Ma-Chen scheme, the length of ciphertext $(c,r,s)$ is the same, i.e., $|p|+2|q|$ bits. For a real system, $p$ and $q$ can be selected as primes with lengths of 1024-bit and 160-bit, respectively. In this setting, the length of ciphertext $(c,r,s)$ is 1344 bits. \vskip 1mm

We now discuss the computation overhead. We only count the numbers of exponentiations that are performed by each party, due to the fact that the exponentiation is the most time-costing computation operation in most cryptosystems. In the Ma-Chen scheme, to generate and verify a ciphertext 3 exponentiations are needed (by Alice and Bob). In our scheme, this number is 5, increased a little. However, to convert a ciphertext $(c,r,s)$ into public verification, Bob does not need to perform any exponentiation in our scheme, while in the Ma-Chen scheme 2 exponentiations are required. In addition, it is the same that the TTP needs to do 2 exponentiations in the public verification procedures of both schemes. In one words, these two schemes have no much difference in the performance. 

\section{Conclusion}

In this paper, we identified two security weaknesses in the Ma-Chen authenticated encryption scheme proposed in \cite{MC03}. Our results showed that their scheme is insecure. Moreover, based on the Schnorr signature scheme, we proposed a new scheme such that all desired security requirements are satisfied.


\begin{thebibliography}{99}




\bibitem{BD98} F. Bao and R.H. Deng, ``A Signcryption Scheme with Signature Directly Verifiable by Public Key,'' {\it Proc. Public Key Cryptography} ({\it PKC'98}), LNCS 1431, Springer-Verlad, pp. 55-59, 1998. 

\bibitem{HMP94} P. Horster, M. Michels, and H. Petersen, ``Authenticated Encryption Scheme with Low Communication Costs,'' {\it Electronics Letters}, Vol. 30, No. 15, pp. 1212-1213, 1994. 

\bibitem{LC95} W.-B. Lee and C.-C. Chang, ``Authenticated Encryption without Using a One Way Function,'' {\it Electronics Letters}, Vol. 31, No. 19, pp. 1656-1657, 1995. 

\bibitem{MC03} C. Ma and K. Chen, ``Publicly Verifiable Authenticated Encryption,'' {\it Electronics Letters}, Vol. 39, No. 3, pp. 281-282, 2003. 


\bibitem{PS00}D. Pointcheval and J. Stern, ``Security Arguments for Digital Signatures and Blind Signatures,'' {\it Journal of Cryptology}, Vol. 13, No. 3, pp. 361-369, 2000. 

\bibitem{Sch91} C. Schnorr, ``Efficient Signature Generation by Smart Cards,'' {\it Journal of Cryptography}, Vol. 4, No. 3, pp. 161-174, 1991.

\bibitem{WLH03} H.-A. Wen, C.-M. Lo, and T. Hwang, ``Publicly Verifiable Authenticated Encryption,'' {\it Electronics Letters}, Vol. 39, No. 19, pp. 1382-1383, 2003.

\bibitem{Zhe97} Y. Zheng, ``Digital Signcryption or How to Achieve Cost (Signature \& Encryption) $<<$ Cost (Signature) + Cost (Encryption),'' {\it Proc. CRYPTO'97}, LNCS 1294, Springer Verlag, pp. 165-179, 1997. 

\bibitem{Zhe97a} Y. Zheng, ``Signcryption and Its Application in Efficient Public Key Solution,'' {\it Proc. Information Security Workshop} ({\it ISW'97}), LNCS 1397, Springer-Verlad, pp. 291-312, 1998. 
\end{thebibliography}
\end{document}